\documentclass{iau}

\usepackage{amsmath}
\usepackage{graphicx}
\usepackage{multirow}

\begin{document}

\lefttitle{Alessa I. Wiggins, et al.}
\righttitle{Evolution of Long-Lived Star Clusters in Galaxy Simulations}

\jnlPage{1}{7}
\jnlDoiYr{2025}
\doival{10.1017/xxxxx}
\volno{395}
\pubYr{2025}
\journaltitle{Stellar populations in the Milky Way and beyond}

\aopheadtitle{Proceedings of the IAU Symposium}
\editors{J. Mel\'endez,  C. Chiappini, R. Schiavon \& M. Trevisan, eds.}

\title{Only the Special Survive: Evolution of Long-Lived Star Clusters in Galaxy Simulations}

\author{Alessa I. Wiggins$^1$, Sarah Loebman$^2$, Peter Frinchaboy$^{1,3}$}
\affiliation{$^1$Department of Physics \& Astronomy, Texas Christian University, Fort Worth, TX 76129, USA}
\affiliation{$^2$Department of Physics, University of California, Merced, 5200 Lake Road, Merced, CA 95343, USA}
\affiliation{$^3$Canada-France-Hawaii Telescope, 65-1238 Mamalahoa Highway, Kamuela, HI 96743, USA}

\begin{abstract}
 In this work, we aim to answer one crucial question behind the discrepancy between chemical trends of field stars and clusters in the Galactic disk: is the chemical gradient mismatch driven by cluster migration and differential survivability as a function of galactic location? To answer this question, we explored the evolution of long-lived ($> 1$ Gyr) star clusters in Milky Way-galaxy simulations. In particular, we investigated why some star clusters remain bound over billions of years.  We have traced the unique trajectories for a sample of open clusters around two FIRE galaxies throughout cosmic time. Additionally, we characterized the small-scale environment surrounding these clusters over their orbital history. We see that clusters across both FIRE galaxies spend the majority of their lives in under-dense regions of gas, except for brief passages where they interact with gas clouds, causing their orbits to be altered. 
\end{abstract}


\maketitle
\section{Introduction}
\indent Open star clusters allow us to trace essential information throughout the rich history of our Galaxy, as we can measure their age and their chemical composition independently. A significant number of long-lived ($> 1$ Gyr) open clusters (OCs) are unexpectedly found at large radii and heights above the Milky Way's disk \citep[][]{CG20}, as compared to young open clusters, and far from the typical disk star-forming regions. In addition to their unusual locations, some of these clusters also exhibit anomalous chemical compositions \citep[e.g.,][]{OCCAM_Myers}.  When we compare the chemical gradient trends with age of field stars to those found in open clusters, a noticeable discrepancy emerges. To investigate whether cluster migration may be behind the observed chemical gradient mismatch for older disk star clusters and field stars, we turn to simulations for further insight; we can accurately trace clusters throughout their history in cosmological simulations, even through significant dynamical changes, that we cannot in the observed cluster population.

\section{Methods}
\indent We examine the evolution of open clusters in galaxies drawn from the \textit{Latte} suite \citep{Latte} of Milky Way-mass zoom-in cosmological simulations generated using the Feedback In Realistic Environments (FIRE)-2 code \citep{FIRE2} --- a high-resolution, physics-based Lagrangian hydrodynamic code, designed to study the formation, evolution, and dynamics of galaxies. FIRE-2 incorporates star formation, stellar feedback, and gas interactions into the modeling process, allowing us to study the impact of the galactic environment on the evolution of clusters in these simulated galaxies.\\
\indent For our analysis, we focus on two Milky Way-like galaxies, \texttt{m12i} and \texttt{m12f}, identifying a combined total of 63 long-lived clusters. These selected clusters are old ($>$ 1 Gyr) and have survived until the present day, making them ideal candidates for comparison against observed Milky Way old open cluster population. In order to understand where open clusters form, if and why they radially migrate, and how they traverse and interact with the Galaxy over time, we track the orbital trajectories of OCs across cosmic time.

\section{Results}
\begin{figure}[t!]
\centering
\includegraphics[scale=0.35]{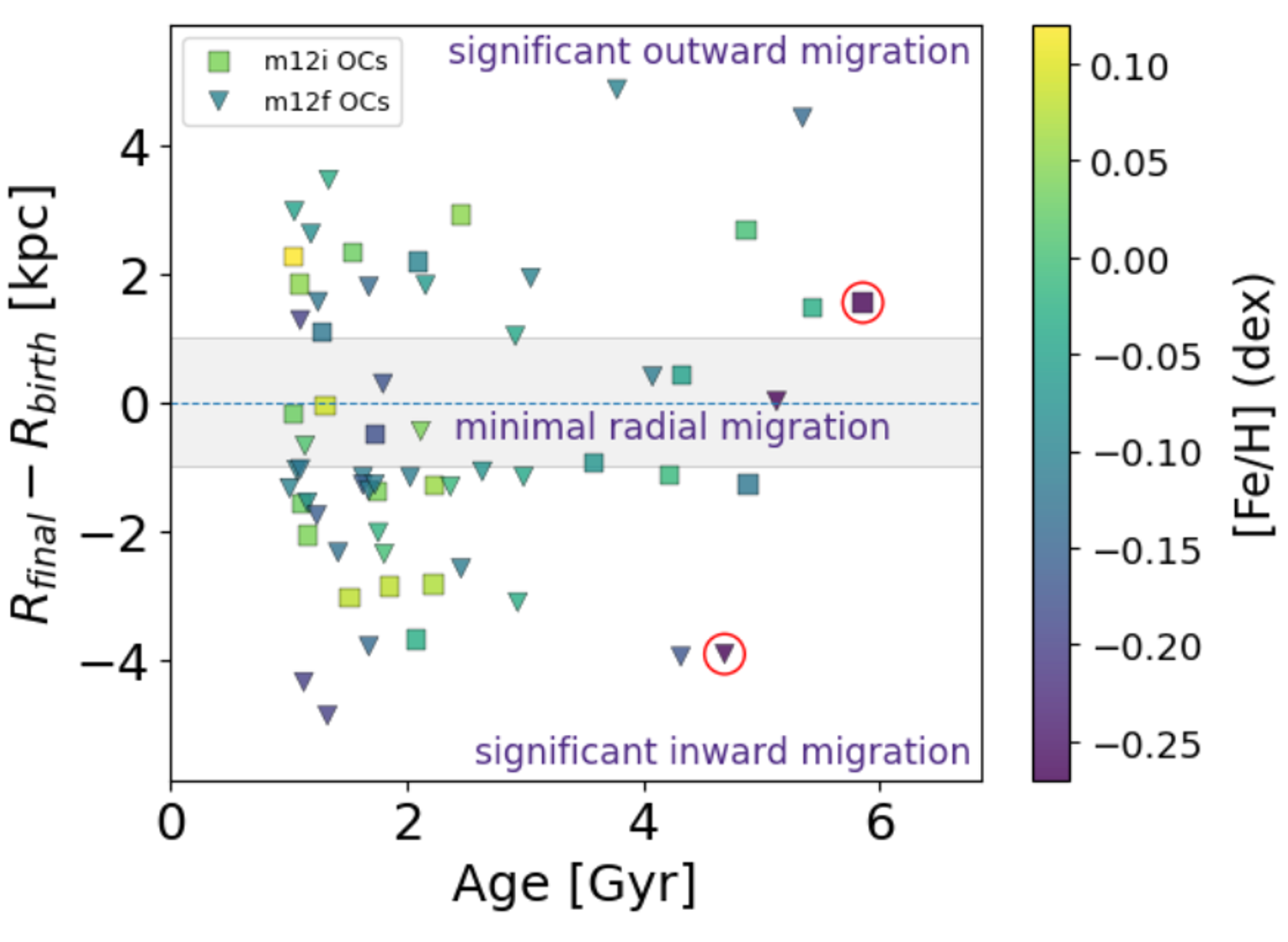}
  \caption{Cluster migration in \texttt{m12i} and \texttt{m12f}: The difference between present-day radius and birth radius for long-lived OCs as a function of cluster age, all color-coded by metallicity. The two circled clustered are examples of inward \& outward migrating clusters (explored in Figure \ref{fig:example_clusters}).}
  \label{figure:cluster_migration}
\end{figure}

\indent We find that clusters do migrate from their location of birth and can survive migration. We observe a diverse range of cluster migration patterns across the two galaxies, with clusters either moving inward, outward, or remaining near their original birth locations (see Figure \ref{figure:cluster_migration}). 

\begin{figure}[t!]
\centering
\includegraphics[scale=0.34]{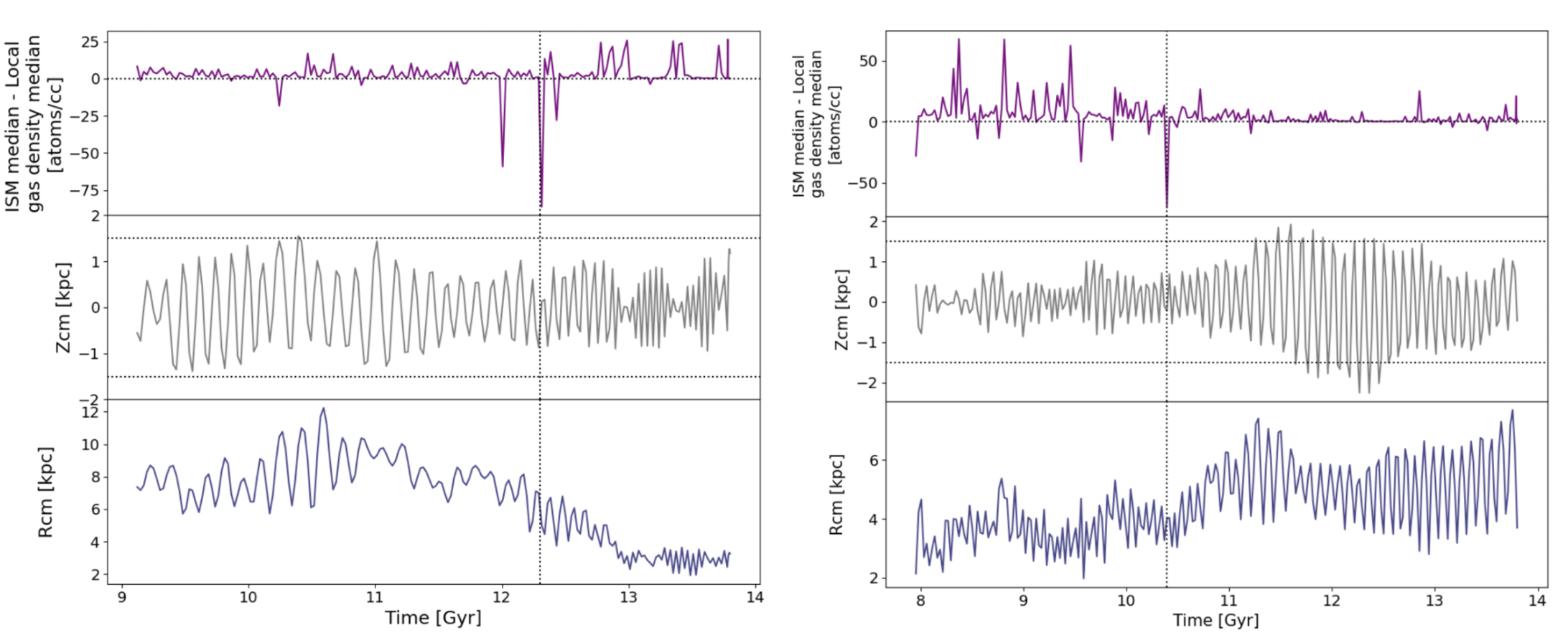} 

  \caption{ 
\textbf{Both clusters undergo major orbital changes.}  
(Left) Inward-moving (aka ``migrating'') 4.69 Gyr cluster selected from \texttt{m12f}.  (Top) The local density relative to the median ISM is typically under-dense except for two brief encounters with dense clouds (the most significant denoted by the vertical line).  (Middle) Cluster's vertical distance from the midplane over time. (Bottom) Cluster's galactocentric radial position over time, which shows that the \textbf{dense interaction drives inward migration and decreases its maximum height above/below the midplane}. (Right) Long-lived (5.86 Gyr) cluster selected from \texttt{m12i} demonstrates outward migration. }
\label{fig:example_clusters} 
\end{figure}

From the full sample, we select two examples of old OCs that survive to present day for further analysis (see Figure \ref{fig:example_clusters}). On the left, we show an illustration of long-lived (4.69 Gyr) inward migrating cluster identified in \texttt{m12f}. The top panel shows the cluster’s local density (100 pc spherical region around the cluster) relative to the global median interstellar medium (ISM) (median gas density in a galactic annulus at the distance of the cluster that is 100 pc wide). We find the cluster typically residing in under-dense regions, except for two brief passages through over-dense clouds. The middle and bottom panels show the cluster’s height and radial position over time, respectively. When the cluster passes through an over-dense region, its orbit is altered. Here, the cluster migrates radially inward, and consequentially its maximum height above the midplane decreases. \\
\indent On the right, we show an example of an older (5.86 Gyr) outward migrating cluster identified in \texttt{m12i}. Similarly, the cluster shown on the left of Figure \ref{fig:example_clusters} generally resides in under-dense regions with one brief passage through a dense gas environment (indicated by the vertical line). After it interacts with the dense environment, the cluster moves radially outward, and consequentially, its maximum height above the midplane increases. Further analysis to the remaining OCs show all of our long-lived clusters systematically occupy under-dense regions of gas throughout their lifetime (Wiggins et al. {\em in prep}).

In the future, we aim to compare the properties of these simulated OCs (e.g., ages and chemical abundances) to their analogs in the observed SDSS-IV/APOGEE \citep[][]{OCCAM_Myers} and SDSS-V/MWM-based OCCAM survey (Otto et al. {\em in prep}).




\vskip-0.1in

\section{Acknowledgments}

We acknowledge funding from the National Science Foundation (AST-2206541, AST-2109234, Aspen Center for Physics PHY-1607611) for the Sloan Digital Sky Survey V has been provided by the Alfred P. Sloan Foundation, the Heising-Simons Foundation, the National Science Foundation, and the Participating Institutions. SDSS acknowledges support and resources from the Center for High-Performance Computing at the University of Utah. The SDSS web site is \href{http://www.sdss.org/}{http://www.sdss.org/}. We also acknowledge the STScI  grant funding (AR-16624).

\end{document}